\documentclass[onecolumn,preprintnumbers,amsmath,amssymb]{revtex4}

\usepackage{graphicx}
\usepackage{subfigure}
\usepackage{dcolumn}
\usepackage{bm}
\usepackage{color}
\usepackage[usenames,dvipsnames]{xcolor}
\usepackage[colorlinks=true,citecolor=Cerulean,linkcolor=RubineRed,urlcolor=Cerulean]{hyperref}

\begin{document}

\title{Focus on Shortcuts to Adiabaticity}%
\author{Adolfo del Campo$^{1,2,3,4}$,  and Kihwan Kim$^{5}$}

\affiliation{$^1$Donostia International Physics Center,  E-20018 San Sebasti\'an, Spain\\
$^2$IKERBASQUE, Basque Foundation for Science, E-48013 Bilbao, Spain\\
$^3$Department of Physics, University of Massachusetts, Boston, MA 02125, USA\\
$^4$Theoretical Division, Los Alamos National Laboratory, Los Alamos, NM 87544, USA\\
$^5$Center for Quantum Information, Institute for Interdisciplinary Information Sciences, Tsinghua University, Beijing, China
}

\date{\today}


%

\date{\today}


\begin{abstract}

Shortcuts to Adiabaticity (STA) constitute driving schemes that provide an alternative to adiabatic protocols to control and guide the dynamics of classical and quantum systems without the requirement of slow driving. Research on STA advances  swiftly with theoretical progress being accompanied by experiments on a wide variety of platforms. We summarize recent developments emphasizing advances reported in this focus issue while providing an outlook with open problems  and prospects for future research. 
\end{abstract}

\maketitle

\section{Introduction}

The understanding of quantum matter far away from equilibrium is a key problem at the frontiers of physics.
Efforts to tailor and control nonadiabatic dynamics have an intrinsic interest and are further motivated by the development of quantum technologies.
In this context, speeding up a given physical process is often desirable. An ubiquitous need is the mitigation of  decoherence and uncontrolled error sources. Other prominent examples include the preparation of ground-state phases of matter in quantum simulators, boosting the power of quantum computers and thermodynamic devices such as batteries, engines, and refrigerators. Ideally, in any application the speedup is to be achieved without sacrificing the efficiency of the process.

Shortcuts to adiabaticity achieve this goal in both classical and quantum systems: they accelerate the evolution of a physical system in a controlled way, stirring the dynamics towards a target state in a nonadiabatic fashion \cite{Chen10}. They provide an alternative to adiabatic techniques in the preparation of a physical system in a given state with high fidelity.
To date, STA have found a broad range of applications; see \cite{Torrontegui2013} for a general review and \cite{Campo2015} with a focus on many-body systems.

\subsection{Novel theoretical approaches}

Among the different approaches to engineer STA, counterdiabatic driving \cite{Demirplak2003,Demirplak2005,Demirplak2008,Berry2009} allows for the engineering of STA in arbitrary physical systems as long as the spectral properties are available.
The focus issue opens with an insightful work by Takahashi \cite{Takahashi_2017} focused on counterdiabatic driving  and  nonequilibrium entropy production. For the latter, a new lower bound is introduced that constitutes a trade-off relation between the duration of the process and a notion of distance to equilibrium. Entropy production is discussed in relation to information geometry \cite{Amari16}, and the Pythagorean theorem under counterdiabatic driving. The analysis of STA in terms of  information geometry  initiated in this work  seems to offer a fruitful perspective deserving  further study.

Patra and Jarzynski have introduced a novel technique to engineer STA using flow fields \cite{Patra_2017}.  They consider a reference adiabatic trajectory for a  system with Hamiltonian $H_0(t)$. Finding the speedup protocol  requires studying the velocity and  acceleration  fields  that  describe  the  time-dependence of the probability density of a given eigenstate of $H_0(t)$.
This approach is particularly welcome as it can be applied to processes lacking dynamical symmetries. It does have the merits of the fast-forward approached developed by Masuda and Nakamura \cite{Masuda2009,Masuda2011} and share its limitation: the protocol varies with the state being driven. This limitation however seems to be broadly present in STA, in the absence of dynamical symmetries. A phase-space understanding of the technique is provided in terms of adiabatic invariants and the method is discussed  in a variety of scenarios including classical, quantum and stochastic systems.

\subsection{Quantum Thermodynamics}

STA have found applications in quantum thermodynamics \cite{Deng13,Campo2014a,Tu14,Beau16,Chotorlishvili16,Abah17,Deng18pra,Deng18Sci}, with important precedents in stochastic thermodynamics \cite{Schmiedl07,Vaikuntanathan08}. In particular, their use has been proposed to suppress quantum friction in finite-time thermodynamics and boost the performance of quantum  engines \cite{Deng13,Campo2014a,Beau16},  see as well preceding work in \cite{Feldmann06}. The feasibility of this goal has recently been demonstrated by the implementation of friction-free superadiabatic strokes \cite{Deng18pra,Deng18Sci}. The manuscript by Diao et al. \cite{Diao_2018} provides an exhaustive account of the state of the art regarding the implementation of STA in Fermi gases and their relevance to finite-time thermodynamics. Superadiabatic strokes are experimentally demonstrated  for both an ideal Fermi gas and a strongly-interacting Fermi gas at unitarity. The latter case is experimentally explored in the low as well as the high-temperature regime, when the evolution of the atomic cloud is governed by viscous hydrodynamics.

Further progress in the engineering of superadiabatic thermal machines has been reported by Li et al. \cite{Li_2018} who theoretically propose the realization of quantum heat engine using a trapped bright soliton as a working substance. The later is kept in a harmonic trap of constant frequency. The thermodynamic cycle consists of a total of four strokes: Work in done onto and by the soliton by ramping up and down nonlinear interactions by through a Feschbach resonance.  These strokes are assisted by STA that ensure the fast control of the shape of the atomic cloud via a modulation of the interaction strength (the reverse control of the interactions by tailoring the shape of the cloud by STA was discussed in \cite{delcampo11epl}).  These strokes are alternated with the coupling to reservoirs of particles to complete the cycle.

\subsection{Cost of Shortcuts to Adiabaticity}

Quantifying the cost of implementing a given protocol (e.g., a STA) is a multi-faceted one with no unique answer.
The implementation of driving protocols in the laboratory can impose constraints on the available  values of a required control field, such as its amplitude, or the rate at which it can be changed.
Auxiliary Hamiltonian terms  used for the control are generally desirable to be local and given by a potential term  (e.g., momentum independent). 
In many-body systems, controls that alter the interactions  are  ideally short-range and few-body.

On theoretical grounds, one can opt for a wide variety of approaches an account,  based on e.g., energetic, thermodynamic, information-theoretic considerations. See, e.g., \cite{Demirplak2008,ChenMuga10,Takahashi_2013,Zheng16,An2016,Coulamy16,Campbell17,Funo17,Bravetti17,Muga17,Calzetta18}.
It should be emphasized that STA designed by counterdiabatic driving can be formulated as time-optimal processes \cite{Takahashi_2013} that satisfy the quantum brachistochrone equation \cite{Carlini05}.

The characterization of the thermodynamic cost of implementing STA, led to the discovery of work-time uncertainty relations in both the quantum \cite{Funo17} and classical \cite{Bravetti17} domain. These results are analogous to the time-energy uncertainty relations and provide tighter bounds to the speed of evolution.  The study by Zhang et al \cite{Zhang_2018}  experimentally demonstrates the validity of these relations in accounting for work fluctuations  along STA.
The experimental platform consists of a cross-shaped superconducting transmon qubit, generally referred to as a Xmon qubit. 
The verified work-time uncertainty relations were derived for counterdiabatic driving and it remains to be elucidated whether similar relations can be found for arbitrary physical processes.

Taking a thermodynamic approach, the usefulness of accounting for the work required to generate a control protocol in comparison to the work extracted has been discussed in \cite{Barato_2017}. In this issue,  Tobalina et al. \cite{Tobalina_2018} argue that accounting for the control system provide a meaningful analysis of the energy consumption of a STA protocol, an analysis that can actually be applied to arbitrary time-dependent processes. Tobalina et al. further  study the cost of  fast and efficient transport of an ion, achieved by modulating the voltages of a segmented Paul trap.

\subsection{Noise-resilient shortcuts}

The implementation of a given protocol in the laboratory often differs from the theoretical prescription due to uncontrolled perturbations and errors in the control fields.  Realistic STA  should therefore be robust. 

Mortensen et al. \cite{Mortensen_2018}  account simultaneously for the cost and robustness of STA  by optimizing a cost functional that incorporates both the resource requirements and a source of perturbations, to favor robustness.  The authors apply this approach to the  $\Lambda$-system, whose control by STA has been well studied since the pioneering works by Demirplak and Rice on counterdiabatic driving \cite{Demirplak2003,Demirplak2005,Chen2010b} and implemented in the laboratory \cite{Du2016}.

The characterization of the cost of STA is intimately related to the dynamics of the system. Techniques to engineer STA in quantum systems remain essentially restricted to isolated systems governed by unitary dynamics. In this context, the coupling to an environment is perceived as a limiting factor of the efficiency of the protocol. Levy et al. \cite{Levy_2018} propose the engineering of  noise-resistant quantum controls by exploiting dynamical invariants. The authors  apply this approach in the presence of dephasing to the  population inversion in a two-level system and the control of coherent and thermal states of a time-dependent harmonic oscillator.

Further  on enhancing the robustness of STA in the presence of noise is reported by Ritland and Rahamani \cite{Ritland_2018}.
The work is motivated by  quantum information processing using  braiding of Majorana zero modes. While these system is  topologically protected against a broad class s of perturbations is affected by high-frequency noise, that limit adiabatic strategies. The authors report the use of simulated-annealing Monte Carlo simulations to design optimal driving protocols that are tailored for a specific noise strength value. While driving protocols take the familiar bang-bang form in the noise-free case, optical protocols are shown to be smooth
in the presence of multiplicative noise.

\section{Progress in specific platforms}

\subsection{Superconducting qubits}

High-fidelity single-qubit quantum gates assisted by STA have been designed and demonstrated in the work reported by Wang et al. using a superconducting Xmon qubit \cite{Wang_2018}. Their theoretical design parallels that proposed in nitrogen-vacancy centers. A method based on the derivative removal by adiabatic gates (DRAG) is employed to create the qubit and decouple its energy levels from excited states.
The authors demonstrate the implementation of unitaries realizing $\pi$ and $\pi/2$ rotations about the X and Z axes as well as a Hadamard gate.
To remove the errors associated with state preparation and readout and characterize the gate fidelity, a Clifford-based randomized benchmarking measurement is used.
The authors report high process and gate fidelities approaching the state-of-the-art values for the considered gates.

A Xmon qubit has been used as well to experimentally verify the work-time uncertainty relations  \cite{Funo17}  in \cite{Zhang_2018}, as already discussed.

\subsection{Trapped ions}

Multiple tasks in trapped-ion quantum technology require the transport of ions.
In this context STA has been proposed as a means to achieve superadiabatic transport. Further experimental progress towards this end has been guided by optical control protocols.

STA may prove useful in heat pump extraction as discussed by Torrontegui et al. \cite{Torrontegui_2018}.
The authors propose the use of inverted harmonic potential to speed up  heat pump realized with a single ion in a tapered trap. The use of  inverted harmonic potentials has been proposed to speed up STA \cite{Chen10} and has been used in the laboratory to control and probe soliton dynamics \cite{Khaykovich02}. In this issue, the authors assess the feasibility of this approach in a trapped ion by analyzing the stability of the scheme  in the presence of micromotion and the finite-temperature of the ion. The effect of noise is minimized by designing low-power protocols.

Further work on the use of STA in trapped-ion systems  has been reported  by Cohn et al. \cite{Cohn_2018}. The authors use a two-dimensional array of trapped ions in a Penning trap that form a Coulomb crystal of $\sim70$  Be$^+$ions,  that is described by the Dicke model. The work focuses on the preparation if entangled states of interest to quantum metrology in a short-time, without the need to rely on  locally-adiabatic state-preparation strategies. The protocols considered are of the bang-bang type: The system is initialized in a product state of a given Hamiltonian, generally simple. An  external parameter is then quenched so that the ensuing dynamics is generated by an intermediate Hamiltonian for a period of time, before to the system is quenched again to a final Hamiltonian of interest (see as well \cite{Shibiao17}). Collective spin observables and the spin distribution are used to experimentally characterize the performance of the protocols in the preparation of the ground-state  in the superradiant phase of the Dicke model.

\subsection{Brownian systems}

In either classical or quantum systems, progress on the engineering of STA  for dissipative processes and in the presence of an environment remains highly limited, see e.g. \cite{Tu14,Vacanti14,Dann18}. 

Chupeau et al. \cite{Chupeau_2018} propose the use of STA to assist the equilibration of a classical Brownian particle in both the underdamped to overdamped regimes. 
Specifically, the particle is trapped in harmonic confinement with time-dependent frequency and remains in contact with a bath that can have a time-dependent temperature. 
A general framework is presented to design the modulation of these two parameters  and  achieve engineered swift equilibration, reaching a prescheduled target state in a given time.

\subsection{Ultracold gases}

Ultracold atoms are the experimental platform first used to demonstrate STA in the laboratory.
STA for expansions  with a thermal atomic cloud \cite{Schaff2010} and a Bose-Einstein condensate \cite{Schaff2011a,Schaff2011} were reported by Labeyrie's group. 
Similar protocols could in principle be used to manipulate arbitrary quantum fluids exhibiting scale invariance \cite{Campo2011,Campo2012b,delcampo2013,Deffner2014}.
A step in this direction was taken by  implementing shorctuts to adiabatic expansions  in a one-dimensional atomic cloud with phase fluctuations \cite{Rohringer2015}.

STA have proved useful  in the manipulation of ultracold gases in processes other than expansions and compressions.
One clear instance is in atom chip technology, where several applications have been  limited by  the long times required to change the location of the the Bose-Einstein condensate cloud. Corgier et al. \cite{Corgier_2018} report the fast and controlled transport of neutral atoms over large distances atom chips, of order 1000 times the atomic cloud size.
The authors further report expansion speeds in the picokelvin regime by tailoring the release and collimation of the atomic cloud. 

The optimization of transport protocols by STA is as well the subject of Ness et al. \cite{Ness_2018}. The authors experimentally demonstrate  shortcuts to adiabatic transport with 
an atomic cloud of cold fermionic potassium atoms  that are confined in an
optical dipole potential created with a Gaussian beam. Previous experiments on STA for transport were restricted to a single particle, e.g., a trapped ion \cite{An2016}. The atoms are at a temperature of $T\approx 300 nK$ and are weakly interacting, in a balanced mixture of Zeeman states with $m=-9/2,-7/2$. Excitations in the final state are probed by measuring the sloshing of the center of mass and assessing the residual excess energy over the ground state.
A variety of STA protocols are considered and experimentally demonstrated, when the  duration of the process duration is of the order of the inverse trapping frequency.

The sped-up  loading of a Bose-Einstein condensate into an optical lattice offers yet another scenario to exploit STA  \cite{Masuda2014}, with  broad applications from quantum simulation to metrology. 
Zhou et al. \cite{Zhou_2018} experimentally demonstrate the  fast transferring of a harmonically-trapped Bose-Einstein condensate, from the ground state to chosen bands of an optical lattice, with high-fidelity. The approach is highly versatile allowing for the population of a single band of either even or odd parity, as well as  the preparation of quantum superpositions between different bands. In addition, it applies to high-dimensional lattice in the presence of degeneracies.  

The dynamics of cold atoms in an optical lattice is as well the subject of the work by Weidner and Anderson \cite{Weidner_2018}, focused on the splitting of an atomic cloud. The authors demonstrated theoretically and experimentally the realization of an atomic beam splitter by tailoring the band-to-band transitions in a shaken optical lattice.  To minimize the heating induced by interatomic interactions, driving frequencies responsible for the phase-modulation of the lattice are kept on resonance with the single- and two-photon transitions between different bands.  The optimization landscape can thus be simplified and restricted to the frequencies of the strongest band and half-band transition resonances. In this simplified landscape, the optimization  is performed via a genetic algorithm. The study is motivated by the prospects of atom interferometry with shaken optical lattices. 

Regarding  theoretical progress, Masuda et al. \cite{Masuda_2018} have as well analyzed the population of specific bands of a periodic potential using the fast-forward technique, building on previous work reported in \cite{Masuda2014}. In addition, their study introduces a novel protocol for phase imprinting,  exploiting a combination of scaling dynamics and the fast-forward technique. The resulting phase imprinting protocol is proposed for the  preparation of highly excited states, describing wave packets with
uniform momentum density.

All these works naturally prompts the question of whether similar control techniques can be applied to strongly-interacting systems, in particular, in ultracold fermions.
 Progress in the use of STA in the strongly-coupled regime has been reported in scenarios assisted by dynamical symmetries (e.g. expansions and compressions of the atomic cloud governed by scale-invariance), using a three-dimensional anisotropic Fermi gas at unitarity \cite{Deng18pra,Deng18Sci}. This includes the realization of superadiabatic expansions and compressions in a high-temperature regime, in which the  evolution can  be described by viscous hydrodynamics \cite{Diao_2018}. Further works are desirable in a general setting, in the absence of dynamical symmetries.

\subsection{Many-body systems: Complexity barrier}

Complex systems with many-degrees of freedom can be expected to be difficult to control.
The complexity barrier is manifest in the application of STA.  Access to the spectral properties of the system become prohibitive, preventing the application of techniques such as counterdiabatic driving. Focusing on the control of a reduced subset of degrees of freedom can pave the way for the use of STA, an ubiquitous strategy in the manipulation of ultracold gases just discussed. This aim is facilitated when the collective mode dynamics is governed exactly or in an approximate fashion by a dynamical symmetry \cite{Chen2010b,Muga09,delcampo11epl,Campo2011,Choi2011,Campo2012b,Jarzynski13,delcampo2013,Deffner2014,Papoular15,Deng18pra,Deng18Sci,Diao_2018}. 

The dynamics of a reduced set of modes can be further controlled when they are weakly coupled to the remaining degrees of freedom. Duncan and del Campo present in \cite{Duncan_2018} the Counterdiabatic Born-Oppenheimer Dynamics (CBOD) as a framework to this end, by harnessing the separation between the energy scales of fast and slow modes to design simplified and efficient STA.

Previous research has shown that even when the spectral properties are available, the required control fields to guide the dynamics of a complex system involves highly nonlocal interactions. This understanding has been gained by analyzing critical quantum spin systems 
 \cite{Campo2012a,Takahashi2013a,Damski2014,Saberi2014,Mukherjee16,Okuyama2016,Sels2017,Takahashi2017}, see \cite{Campo2015} for a review.

Suppressing excitations in many-body systems is manifestly challenging across a second-order or quantum  phase transition, as the relaxation time diverges in the neighborhood of the critical point. This well-known phenomenon known as critical slowing down thus prevent the application of adiabatic strategies in this context. This has important implications in the preparation of ground-state phases of matter in quantum simulators. The Kibble-Zurek mechanism provides a convenient framework to analyze the breakdown of adiabaticity and predicts that the density of excitations scales as a universal power law on the time scale in which the transition is crossed \cite{Campo2014b}.

 In the crossing of a phase transition,  local driving  by spatially inhomogeneous fields can pave the way for defect suppression whenever the breaking of symmetry on a part of the system can subsequently bias the dynamics in the neighboring regions. The role of causality to assist adiabaticity in quantum systems was recognized in \cite{ZurekDorner08}. Building on a large body of literature in both classical and quantum systems, the Inhomgeneous Kibble-Zurek mechanism has been formulated \cite{del_Campo_2013}. 
 
Local driving  by a control field with a single inhomogeneous front has been recently discuss to favor adiabatic dynamics in disordered spin chains \cite{Rams_2016}, an approach extended in 
Mohseni et al. \cite{Mohseni_2018} by considering the use of multiple fronts. The authors provide a framework to analyze the ensuing quantum critical dynamics
 using strong-disorder renormalization group (SDRG), and characterize the  scaling of the gaps with the system size,  that is improved with respect to the  standard adiabatic evolution in homogeneous systems.

In the manipulation of many-body spin systems, other kinds of local driving  may be of interest. Pyshkin et al. \cite{Pyshkin_2018} consider operations in which interactions between spins are completely switched off (``cutting'') or raised from zero value (``stitching''). Such operations  change the topology of a spin system and the authors analyze theoretically their high-fidelity nonadiabatic implementation by relying exclusively on local controls.  Cutting and stitching of spin chains  can be considered a kind of local quench protocols to study nonequilibrium dynamics and the spreading of correlations. They are as well of interest for the control and removal of impurities and the description of the coupling of a system to a bath, e.g.,  in quantum thermodynamics.

\subsection{Multi-partite systems: Entanglement and quantum correlations}

The development of STA to control quantum correlations in  multi-partite systems constitutes an important application in which progress  has somewhat restricted.
Hatomura \cite{Hatomura_2018} addresses the preparation of cat-states in bosonic Josephson junctions utilizing  STA techniques. The analysis put forward relies on the description of the bosonic Josephson junction  in terms of  the semiclassical limit of the Lipkin-Meshkov-Glick model with a collective spin  degree of freedom. The counterdiabatic driving of this system  had previously been explored in \cite{Campbell15}. The Holstein-Primakoff transformation is used to show that the preparation of  cat-states involves 
changing the energy landscape from an effective single-well to a double-well structure.
Across a quantum phase transition, the magnitude of counterdiabatic fields  is known to diverge with the system size \cite{Campo2012a}, a difficulty that is circumvented by explicitly taking into account the finite size of the system.

STA can be applied to the generation of other entangled states, including cluster states. Kyaw and Kwek \cite{Kyaw_2018} show the feasibility of preparing one-dimensional cluster states by analyzing the counterdiabatic driving of the one-dimensional Kitaev honeycomb model.  This model belongs to the family of quasi-free fermions in which the exact counterdiabatic form can be found \cite{Campo2012a}. While the resulting protocols enhance the fidelity of cluster state preparation, they require multiple-body interactions the realization of which is generally a challenging task. Ideally, this work would motivate similar studies in higher dimensional settings where cluster states are a resource for measurement-based quantum computation.

\section{Summary and Conclusions}

This Focus issue reflects the pace of research on STA with an attractive balance between theoretical and experimental progress.
As a control tool, STA can be expected to have manifold applications and 
one can anticipate further developments arising from an interdisciplinary approach,  drawing ideas from different fields. Research on STA is likely to continue  combining  atomic and molecular physics, quantum information, ultracold quantum matter and optimal control. Exciting prospects arise in the combination with quantum computation and annealing, quantum communications, quantum chemistry, information geometry and machine learning, among other examples. On the classical domain, applications can be envisioned, e.g.,  in mechanical engineering, stochastic systems, soft-condensed matter and plasma physics. We hope that the collection of articles reported in this issue encourages the reader to pursue further research on the topic.

\section*{Acknowledgements}

This research is supported by  the John Templeton Foundation and the National Key Research and Development Program of China under Grants No.~2016YFA0301900, No.~2016YFA0301901, No.~2017YFA0303900, and  No.~2017YFA0304004, and the National Natural Science Foundation of China Grants No.~11574002.

\bibliographystyle{iopart-num}

\bibliography{NJPFocusOnSTA}

\end{document}